\documentclass[letterpaper,twocolumn,english,aps,prb,floatfix,showpacs,amsfonts]{revtex4}
\usepackage[T1]{fontenc}
\usepackage[latin1]{inputenc}
\usepackage{amsmath}
\usepackage{babel}
\usepackage{graphics}
\usepackage{amssymb}
\usepackage{bm}



\begin{document}
\title{The correlation exponent $K_\rho$ of the one-dimensional Kondo lattice model}
\author{J. C. Xavier}
\affiliation{Instituto de F\'{\i}sica Gleb Wataghin, Unicamp,
C.P. 6165, 13083-970 Campinas SP, Brazil}
\author{E.~Miranda}
\affiliation{Instituto de F\'{\i}sica Gleb Wataghin, Unicamp, C.P. 6165,
13083-970 Campinas SP, Brazil}
\date{\today}
\begin{abstract}
We present results for the correlation exponent $K_\rho$ of the
Tomonaga-Luttinger liquid description of the one-dimensional Kondo
lattice as a function of conduction electron density and coupling
constant. It is obtained from the first derivative of the Fourier
transform of the charge-charge correlation function. We also show that
the spin correlation function can only be described in this picture if
we include logarithmic corrections, a feature that had been previously
overlooked. A consistent description of both charge and spin sectors
is then obtained.  Finally, we show evidence that the spin sector of
the dimerized phase at quarter-filling is gapless.
\end{abstract}
%75.10.-b General theory and models of magnetic ordering
%71.10.Pm Fermions in reduced dimensions (anyons, composite fermions,
%Luttinger liquid, etc.)
%73.61.Ph Polymers; organic compounds
%72.80.Le Polymers; organic compounds (including organic semiconductors)
%75.30.Mb Valence fluctuation, Kondo lattice, and heavy-fermion
 %phenomena
%71.30.+h Metal-insulator transitions and other electronic transitions

\maketitle
The Kondo lattice model is the simplest model believed to describe the
low energy physics of heavy fermion materials.\cite{livro} Its
one-dimensional version has been thoroughly studied in the last 10
years and a great deal of understanding has been gained. However, some
outstanding issues remain, some of which may have implications in the
higher dimensional cases. For example, the question of whether the
localized spins should be counted in a Luttinger's theorem
determination of the size of the Fermi momentum is still
controversial.\cite{yamanakaetal,japas1,japas2,small,pivovarovsi}
Furthermore, even the phase diagram still presents some novel
surprising phases: at quarter conduction electron filling the spins
are dimerized and the charge sector is gapped.\cite{dimer} The latter
phase may be at the origin of the spin-Peierls phase observed in the
quasi-one-dimensional organic compounds (Per)$_{2}$M(mnt)$_{2}$ (M=Pt,
Pd).\cite{org} At a \emph{generic incommensurate filling} however, the
system is gapless in both the spin and charge sectors and it is
reasonable to assume\cite{tllnote} that it is a Tomonaga-Luttinger
liquid (TLL).\cite{voit} In this paper, we will assume that this is
the case. In an attempt to systematically characterize this behavior,
we have determined the non-universal TLL exponent $K_\rho$ as a
function of coupling constant and conduction electron density. We
found that a consistent picture of charge and spin sectors can be
obtained, only if logarithmic corrections are included in the spin
correlations.  Moreover, we also show that the spin excitation
spectrum of the quarter-filled case is gapless. We give arguments
showing that the presence of dimerization and the absence of a spin
gap are not mutually exclusive.

We considered the one-dimensional spin-\(\frac{1}{2} \) Kondo lattice
Hamiltonian with \( L \) sites
\[
H=-\sum ^{L-1}_{\substack{j=1 \\ \sigma=\pm 1} }
c^{\dagger }_{j,\sigma }c^{\phantom
{\dagger }}_{j+1,\sigma }+h. c.+
J\sum^{L}_{j=1}\mathbf{S}_{j}\cdot \mathbf{s}_{j},\]
 where \( c_{j\sigma } \) annihilates a conduction electron in site
\( j \) with spin projection \( \sigma/2  \), \( \mathbf{S}_{j} \)
is a localized spin-\( \frac{1}{2} \) operator and \(
\mathbf{s}_{j}= \frac{1}{2} \sum _{\alpha \beta }c^{\dagger
}_{j,\alpha }\bm {\sigma }_{\alpha \beta }c^{\phantom {\dagger
}}_{j,\beta } \)
is the conduction electron spin density operator. \( J>0 \) is the
Kondo coupling constant between the conduction electrons and the local
moments and the hopping amplitude has been set to unity to fix the
energy scale. We studied the model with the density matrix
renormalization group (DMRG) technique \cite{white} with open boundary
conditions. We used the finite-size algorithm for sizes up to \( L=120
\) keeping up to \( m=600 \) states per block.  The discarded weight
was typically about \( 10^{-5}-10^{-8} \) in the final sweep.

TLL's  with periodic boundary conditions and SU(2) symmetry have
charge and spin correlation functions given asymptotically
by\cite{voit}
\begin{eqnarray}
\left\langle \delta n\left( 0\right) \delta n\left( x\right)
\right\rangle  & =
& \frac{K_{\rho }}{\left( \pi x\right) ^{2}}+A_{1}\frac{\cos
\left( 2k_{F}x\right) }{x^{K_{\rho }+1}}\nonumber \\
 & + & A_{2}\frac{\cos \left( 4k_{F}x\right) }{x^{4K_{\rho }}},
\label{nn-corr} \\
\left\langle \mathbf{S}^{T}\left( 0\right) \cdot
\mathbf{S}^{T}\left( x\right) \right\rangle  & = & \frac{1}{\left(
\pi x\right) ^{2}}+B_{1} \frac{\cos \left( 2k_{F}x\right)
}{x^{K_{\rho }+1}},\label{ss-corr}
\end{eqnarray}
where $\delta n\left( x\right)=n(x)-\left\langle n(x) \right\rangle$,
\( \mathbf{S}^{T}(j)=\mathbf{S}_{j}+\mathbf{s}_{j} \), \( K_{\rho } \)
is the non-universal charge correlation exponent and
\( k_{F} \) is the Fermi momentum. Local charge perturbations, such as
introduced by impurities or boundaries, induce density oscillations,
called Friedel oscillations. In the case of a TLL, they take the
form
\cite{varios,japas1,japas2}

\begin{equation}
\label{chargeFriedel}
\left\langle \delta n\left( x\right) \right\rangle  =
C_{1}\frac{\cos \left(
 2k_{F}x\right) }{x^{\left( K_{\rho }+1\right) /2}}+
C_{2}\frac{\cos \left( 4k_{F
}x\right) }{x^{2K_{\rho }}}.
\end{equation}

The main goal of this work is to present \( K_{\rho } \) as a function
of the conduction electron density and the Kondo coupling $J$ for the
one-dimensional Kondo lattice model. A previous work\cite{japas2}
determined \( K_{\rho } \), but only for the density $n=2/3$.
Besides, in that work, the authors argued that the system has a
``large'' Fermi surface, with $2k_F^L=\pi(n+1)\pmod{2\pi}$, not a
``small'' one with $2k_F^S=\pi n\pmod{2\pi}$. Indeed, under some
assumptions, the presence of low-lying excitations with momentum
$2k_F^L$ can be proved.\cite{yamanakaetal} Assuming a ``large'' Fermi
surface, the numerical results show that the dominant term in the
charge Friedel oscillations is the second one in
Eq.~(\ref{chargeFriedel}). From the decay of the envelope function of
this term, $K_\rho$ was determined at $n=2/3$.\cite{japas2} However,
more recent work has called into question the presence of a ``large''
Fermi surface, particularly for small $J$.\cite{small} If the Fermi
surface is small, both terms in Eq.~(\ref{chargeFriedel}) oscillate
with the same period at $n=2/3$ and the envelope function method
cannot be unambiguously applied.  In order to avoid this ambiguity, we
determined \( K_{\rho } \) from the first term in Eq.~(\ref{nn-corr}),
or equivalently, from the derivative of the Fourier Transform of the
charge-charge correlation function at $q=0$
\begin{equation}
K_{\rho }=\pi \left. \frac{ \partial C(q) }{\partial q}\right|_{q=0},
\label{krho}
\end{equation}
where
$$
 C(q)=\frac{1}{L}\sum _{j,k}e^{iq\left( j-k\right) }
\left\langle \delta n\left( j\right) \delta n\left( k\right)
\right\rangle .
$$
This method has been shown to give very accurate results by Daul and
Noack.\cite{daulnoack} These authors determined the exponent \(
K_{\rho } \) (by the DMRG technique) for the one-dimensional Hubbard
model and found good agreement with the exact results. For this
reason, in the present work we will use this procedure to estimate the
exponent \( K_{\rho } \).

%######################################################################
\begin{figure}
{\centering\resizebox*{3.2in}{!}{\includegraphics{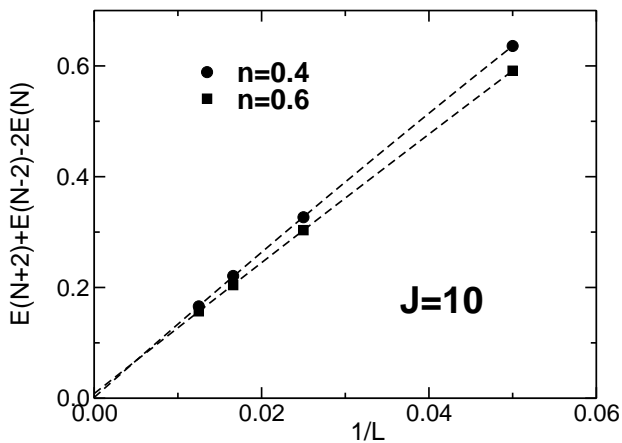}} \par}
\caption{ \label{fig1} Charge gap vs $1/L$ for $J=10$ and densities
$n=0.4$ and $n=0.6$. Here, $E(N)$ is the ground state energy of the
sector with $N$ electrons.  The dashed lines are fits to
$\Delta_{\infty}+c_1/L+c_2/L^2$.  }
\end{figure}
%######################################################################

Haldane has conjectured that the TLL is the generic universality class
of one-dimensional gapless systems.\cite{haldconj} Although a rigorous
proof usually relies on the integrability of the model,
renormalization group arguments confirm this conjecture in
paramagnetic phases.\cite{voit} Much less is known about the case of
systems with ferromagnetic ground states.\cite{bartoschetal} However,
even in this case, the spin sector usually decouples from the charge
sector and it is possible for the latter to remain a TLL. The
one-dimensional Kondo lattice model is ferromagnetic for sufficiently
large $J$.\cite{rkondo} In Fig.~\ref{fig1} we show its charge gap as
a function of system size for $J=10$ (inside the ferromagnetic phase)
and the densities $n=0.4$ and $n=0.6$. The extrapolated values suggest
that the ferromagnetic phase, like the paramagnetic one, has no charge
gap.  Thus, it is quite natural to expect that, inside the
ferromagnetic phase, the charge sector may also be described as a TLL
and we will assume this to be the case.  In fact, as we will see
below, our results are consistent with this assumption. Note that this
appears to happen also in the case of the Hubbard model with
next-nearest neighbor hopping.\cite{daulnoack}

%######################################################################
\begin{figure}
{\centering\resizebox*{3.0in}{!}{\includegraphics{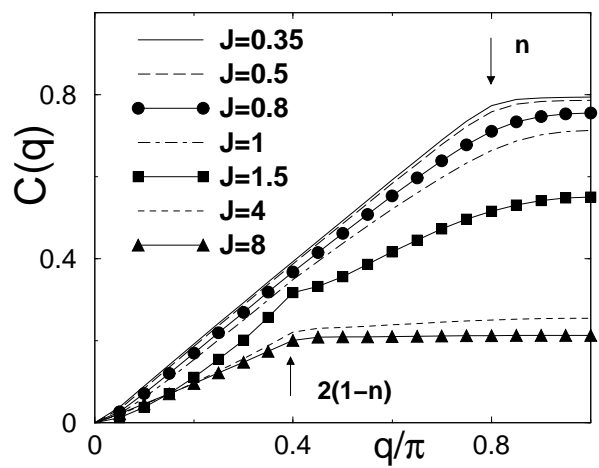}} \par}
\caption{
\label{fig2}
Fourier transform $C(q)$ versus momentum for severals values of  $J$,
$L=40$, and density $n=0.8$. The arrows indicate the position of the
cusp.
 }
\end{figure}
%######################################################################

We first focus on the general $q$-dependence of $C(q)$.  In
Fig.~\ref{fig2} we present the Fourier transform of the charge-charge
correlation function for $n=0.8$, $L=40$, and several values of
$J$. We have checked that the \emph{qualitative} behavior of $C(q)$
presents no finite size effects, and also observed that the simple sum
rule $C(0)=0$ is satisfied (within the accuracy of the DMRG) for all
values of density and Kondo coupling $J$ shown. For small values of
$J$ and all densities, $C(q)$ increases linearly with $q$ up to $q=\pi
n$, and then saturates at $C(q)=n$ for $n<q/\pi<1$. On the other hand,
for large Kondo coupling $C(q)$ increases linearly with $q$ up to
$q=2\pi n \pmod{2\pi} <\pi$ and then saturates at $C(q)=n$ ($1-n$) for
$n<1/2$ ($n>1/2$) and $2\pi n \pmod{2\pi}<q<\pi$.

In order to get some insight into the behavior of $C(q)$ we consider
free fermions with spin-S in a one dimensional nearest-neighbor
tight-binding lattice. In this case, the Fourier Transform of the
charge-charge correlation function $C_0^S(q)$ is
\begin{eqnarray}
 \frac{C_0^S(q)}{(2S+1)}=\left\{
\begin{array}{c}
{q/2\pi} \hspace{.3cm} 0\le q/\pi \le 2 m, \\
m \hspace{.3cm} 2m \le q/\pi \le 1,
\end{array}
\right.
\label{cqs}
\end{eqnarray}
where $m=\min\left[n/(2S+1),1-n/(2S+1)\right]$. We will need two particular
cases, with the restriction $n<1$. For spin-$\frac{1}{2}$ fermions
\begin{equation}
C_0^{1/2}(q)=\left\{
\begin{array}{c}
{q/\pi} \hspace{.3cm} 0\le q/\pi \le n, \\
n \hspace{.3cm} n \le q/\pi \le 1,
\end{array}
\right.
\label{cqhalf}
\end{equation}
while for spinless fermions, if $n<1/2$,
\begin{equation}
C_0^{0}(q)=\left\{
\begin{array}{c}
{q/2\pi} \hspace{.3cm} 0\le q/\pi \le 2n, \\
n \hspace{.3cm} 2n \le q/\pi \le 1, 
\end{array}
\right.
\label{cq0a}
\end{equation}
and if $n>1/2$,
\begin{equation}
C_0^{0}(q)=\left\{
\begin{array}{c}
{q/2\pi} \hspace{.3cm} 0\le q/\pi \le 2(1-n), \\
1-n \hspace{.3cm} 2(1-n) \le q/\pi \le 1.
\end{array}
\right.
\label{cq0b}
\end{equation}

Our results for $C(q)$ in the one-dimensional Kondo lattice model all
tend to the free spin-$\frac{1}{2}$ case when $J\rightarrow 0$ (see
Fig.~\ref{fig2}), as expected. Besides, for $J\gg 0$, $C(q)$ tends to
the $S=0$ case, $C_0^0(q)$. This is also to be expected, since in this
case the conduction electrons form unbreakable mobile singlets with
the localized spins, effectively behaving like spinless
fermions.\cite{rkondo}

%######################################################################
\begin{figure}
{\centering\resizebox*{3.2in}{!}{\includegraphics{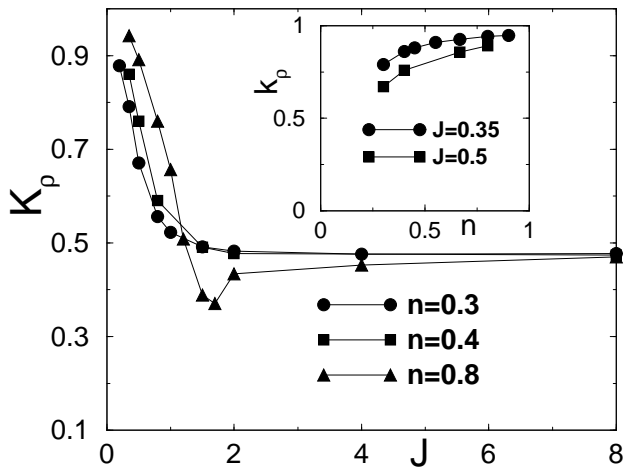}} \par}
\caption{ \label{fig3} The exponent  \( K_{\rho }\)
obtained with Eq. \ref{krho} as a function of $J$ with $L=40$. The
densities are indicated. Inset:  \( K_{\rho }\) vs density for
J=0.35 and J=0.5.
% In all cases $L=40$, expect for the density
% $n=2/3$ where $L=42$.
 }
\end{figure}
%######################################################################

The cusp of $C(q)$ at $q=2k_{F}^{S}=n\pi$, for small values of $J$ is
the signature that the charge density oscillations (Eq.~\ref{nn-corr})
are dominated by the $2k_{F}^{S}$ term. As we increase $J$ the cusp
moves to $q=2\pi n \pmod{2\pi}$. At first sight, this might seem like
an indication that the system crosses over from a $2k^S_{F}$-dominated
region to a $4k^S_{F}$-dominated one as $J$ increases. Indeed, this is
what happens in the Hubbard model when we increase the on-site
repulsion $U$.\cite{rai} However, since $4k_F^S=2\pi n =
4k_F^L\pmod{2\pi}$, the change might be due to a phase transition from
a small Fermi surface to a large one.\cite{pivovarovsi} Indeed, there
have been indications of an intervening ferromagnetic phase at
intermediate values of $J$,\cite{phaseferr} which could give rise to
this change. Unfortunately, the study of other (spin-spin) correlation
functions has not shed any light on the issue:\cite{small} the size of
the Fermi surface for intermediate values of $J$ remains an open
question.

In Fig.~\ref{fig3} we show the exponent \( K_{\rho } \) calculated
through Eq.~(\ref{krho}) for several values of $n$ and Kondo coupling
$J$. For all densities we see that \( K_{\rho } \) tends to unity when
$J\rightarrow 0$, in agreement with our expectation that the system
tends to a non-interacting spin-$\frac{1}{2}$ electron gas with \(
K_{\rho }^{0}=1 \). On the other hand, in the strong coupling limit $
K_{\rho } \sim 0.5$, as expected for free spinless fermions
(cf. Eqs.~\ref{krho} and \ref{cq0a}).  We have observed that for
densities $n>1/2$ and $J\sim 1.5$, \( K_{\rho } \) attains its
smallest values. This is a region where charge oscillations are
enhanced (see, e. g., Fig.~1 of Ref.~\onlinecite{japas1}).  Actually,
for some values of $J$ and $n>1/2$ we were unable to determine
\(K_{\rho } \), as for example at $J=1.7$ and $n=0.9$. At these
points, our data did not satisfy the simple sum rule $C(0)=0$ even
increasing the truncation $m$ up to $m=600$. The exponent \( K_{\rho }
\) also did not converge as a function of $m$. A truncation of $m=600$
is more than enough to get precise values in other parameter regions.
Typically, for small values of $J$ and $L=40$, \( K_{\rho } \)
obtained with truncations $m=400$ and $m=600$ differ by less than
$5. 10^{-2}$ and $C(0)\sim 10^{-4}$ with $m=400$. It is interesting to
note that this region where the charge oscillations are strongest
corresponds to the ferromagnetic phase at intermediate
$J$.\cite{phaseferr} For completeness, we also show in the inset of
Fig.~\ref{fig3} the dependence of \( K_{\rho } \) on the density for
$J=0.35$ and $J=0.5$ with $L=40$. As we can see, \( K_{\rho } \)
decreases with increasing $J$.

%######################################################################
\begin{figure}
{\centering\resizebox*{3.0in}{!}{\includegraphics{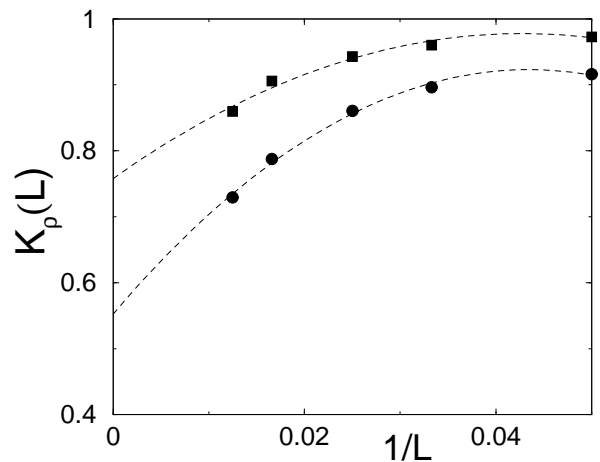}}
\par} \caption{ \label{fig4}
 The exponent \( K_{\rho }\) vs $1/L$ for $J=0.35$,
 $n=0.4$ (circles) and $n=0.8$ (squares).  The dashed lines are fits
 according to Eq. \ref{krhol}, with $K_{\rho }=0.55$ (0.76) $a=17.0$
 (10.3) and $b=-196.1$ (-120.5) for $n=0.4$ (0.8).  }
\end{figure}
%######################################################################

Given this qualitative behavior of \( K_{\rho } \) as a function of
$J$, we now set out to determine some quantitative values of the
exponent. For this, we must be careful to take into account
finite-size effects. In Fig.~\ref{fig4} we show \( K_{\rho } \) for
the densities $n=0.4$ (circles) and $n=0.8$ (squares) as a function of
$1/L$ at $J=0.35$. In order to incorporate the finite-size dependence,
we determined the extrapolated exponent assuming that \( K_{\rho }(L)
\) behaves like
\begin{equation}
 K_{\rho }(L)=K_{\rho }+a/L + b/L^2.
\label{krhol}
\end{equation}
In Fig.~\ref{fig4} the dashed lines are fits to our data using Eq.
\ref{krhol}. The exponents  \( K_{\rho } \) obtained through the fits are
$0.55$ and $0.76$ for $n=0.4$ and $n=0.8$, respectively. The values of
\( K_{\rho } \) shown in the inset of Fig.~\ref{fig3} for small values
of $J$ and $L=40$ should be seen as upper limits. From the uncertainty
of about 0.05 in the values of \( K_{\rho } \) for fixed $L$, we
estimate the error in the extrapolated values to be $\alt 0.1$.

%######################################################################
\begin{figure}
{\centering\resizebox*{3.0in}{!}{\includegraphics{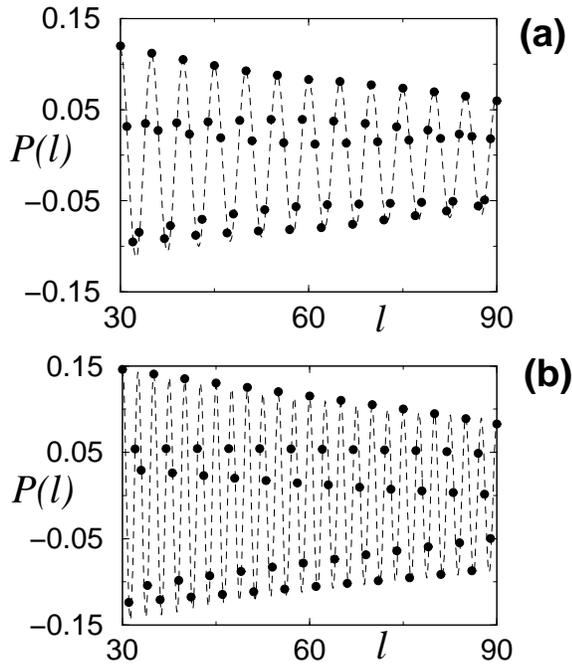}} \par}
\caption{ \label{fig5} 
The spin-spin correlation function for
densities $n=0.4$ (a) and $n=0.8$ (b) ($J=0.35$ and $L=120$).
The dashed line is a fit of Eq. \ref{ss-corr-log} with
$\alpha=4.0$ and $ K_{\rho }=0.55$ for $n=0.4$, and $\alpha=5.1$
and $ K_{\rho }=0.76$ for  $n=0.8$.
 }
\end{figure}
%######################################################################

Now that we have accurately determined the exponent \( K_{\rho }
\) we should be able to describe all correlation functions, since
the only parameters needed are \( K_{\rho } \) and the Fermi momentum
$k_F$ (\(K_\sigma=1\) because of SU(2) symmetry). In particular, we
can cross-check our results with the spin-spin correlation function
(Eq.~\ref{ss-corr}).  Previous work showed that for small values of
the Kondo coupling $J$ the size of the Fermi surface is
small,\cite{small} so that $k_F$ is fixed (see also
Ref. \onlinecite{phaseferr}).  To eliminate the effect of the open
boundaries on the spin-spin correlation function, we considered a
large system and averaged the correlations over pairs of sites
separated by a given distance $j$ to get $\left\langle
\mathbf{S}^{T}\left( 0\right) \cdot
\mathbf{S}^{T}\left( l\right) \right\rangle$, as discussed by
other authors.\cite{small,daulnoack,whitex} In Fig.~\ref{fig5},
the circles correspond to  this averaged spin-spin correlation
function $P(l)=\left\langle \mathbf{S}^T\left( 0\right) \cdot
\mathbf{S}^T\left( l\right) \right\rangle_{aveg}$ for $n=0.4$ and
$n=0.8$ at $J=0.35$ and $L=120$. We restricted the values to the
interval $30\le l\le 90$, because the TLL description only makes
sense asymptotically and large values of $l$ may be compromised by the
open boundaries. \emph{A direct attempt at fitting the data of
Fig.~\ref{fig5} with Eq.~\ref{ss-corr} yields \( K_{\rho } < 0 \),
which is clearly incorrect.} We believe the discrepancy is due to
logarithmic corrections, which are well established in other models
with SU(2) symmetry, e. g. the Heisenberg
model.\cite{giamarchi,hallberg} Thus, assuming a generic form
\begin{eqnarray}
\left\langle \mathbf{S}^{T}\left( 0\right) \cdot
\mathbf{S}^{T}\left( x\right) \right\rangle  & = & \frac{1}{\left(
\pi x\right) ^{2}}+B_{1} \frac{\cos \left(
2k_{F}x\right)\ln^{\alpha}{x} } {x^{K_{\rho
}+1}},\label{ss-corr-log}
\end{eqnarray}
we can produce an excellent fit to the numerical results \emph{using
the exponents $K_{\rho }$ independently obtained before} and only two
fitting parameters: $\alpha=4.0$ ($\alpha=5.1$) and $B_1=0.17$
($B_1=0.11$) for $n=0.4$ ($n=0.8$).  This is seen as the dashed line
in Fig.~\ref{fig5}.  Logarithmic corrections thus appear to be crucial
for a complete description of the spin correlations of the
one-dimensional Kondo lattice model. As far as we know, this point has
not been stressed before. We note that the above values of $\alpha$
differ from the expected values of $1/2$.\cite{affleckpc} A more
accurate determination of the exponent of the log correction may
require much larger system sizes.

%######################################################################
\begin{figure}
{\centering\resizebox*{3.0in}{!}{\includegraphics{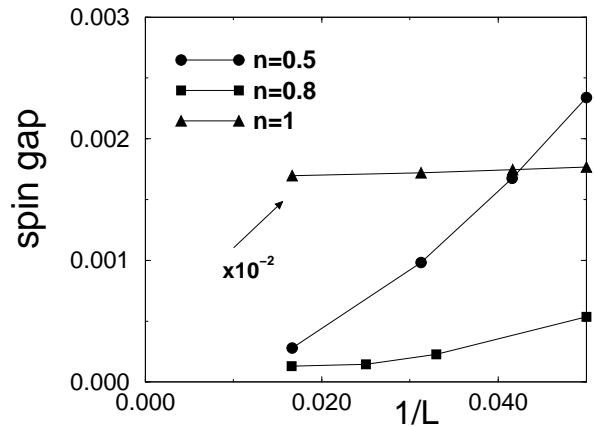}}
\par} \caption{ \label{fig6} Spin gap as function of $1/L$ for the
densities $n=0.5$, $n=0.8$ and $n=1$ ($J=1.2$). The data for $n=1$
have been multiplied by $10^{-2}$ for comparison.
 }
\end{figure}
%######################################################################

Finally, we would like to address the quarter-filled case, $n=1/2$,
which has been shown to exhibit spin dimerization.\cite{dimer} At this
filling, the system has a charge gap and the charge sector cannot be
described as a TLL.\cite{dimer} Furthermore, since the spins are
dimerized, we would naively expect a finite spin gap, as in the
frustrated $J_1-J_2$ Heisenberg model.\cite{haldanedimer} We would now
like to show that in fact the spin sector is gapless.  In
Fig.~\ref{fig6} we show the spin gap as a function of the lattice size
$L$ for $J=1.2$ and densities $n=0.5$, $n=0.8$, and $n=1.0$. We chose
to work with a large $J$ value in order to produce large spin gaps,
since the spin gap generally increases with $J$.  At half-filling,
where the system is known to be fully gapped,\cite{rkondo} the data
clearly tend to saturate at a non-zero value in the thermodynamic
limit. By contrast, at $n=0.5$ and $n=0.8$ the data strongly indicate
that the spin sector is gapless. Thus, the dimerized phase discovered
in Ref.~\onlinecite{dimer} has a charge gap but no spin gap.  In that
reference, the effect of dimerization in the localized spin sub-system
on the conduction electrons was discussed. If we integrate out the
local moments, an effective exchange interaction among the conduction
electrons is generated. This is in a sense the ``complement'' of the
RKKY interaction, which induces an effective exchange interaction
between local moments once the conduction electrons are integrated
out. This effective exchange interaction is proportional to the static 
spin susceptibility of the dimerized localized spins
\begin{equation}
H_{eff} \sim J^2 \sum_{jk} \chi^l (j-k) \mathbf{s}_{j} \cdot \mathbf{s}_{k}.
\end{equation}
If only nearest neighbor terms are retained \(\chi^l
(j-k)=\delta_{j,k+1}D(j)\), where \(D(j)\sim (-1)^j D_0\) is the dimer
order parameter. This leads to a \emph{staggered exchange interaction}
between conduction electron spin densities
\begin{equation}
H_{eff} \sim \sum_{j} (-1)^j \mathbf{s}_{j} \cdot
\mathbf{s}_{j+1}.
\label{effexch}
\end{equation}
This kind of interaction can be analyzed through
bosonization.\cite{voit} Among the many terms that are generated,
those which involve combinations like
\(e^{-i4k_Fx}\psi^\dagger_R(x)\psi_L(x)\psi^\dagger_R(x+1)\psi_L(x+1)\) 
will have just the right oscillating factor to cancel the $(-1)^j$ in
Eq.~\ref{effexch}, \emph{since \(4k_F=\pi\) at quarter filling}. One
of the terms is \(\sim \sin(2\sqrt{2}\phi_\rho)\) (in the notation of
Ref.~\onlinecite{voit}), which is relevant and opens a charge gap if
\(K_\rho<1\).\cite{voit} The DMRG results show that this condition
is fulfilled throughout the phase diagram. Thus, the above analysis
seems to be consistent with the presence of a charge gap. A term of
the form \(\sim \sin(2\sqrt{2}\phi_\sigma)\) is also
generated. However, it is \emph{marginal} if \(K_\sigma=1\) and only
generates a spin gap if the coefficient has the right sign. We
conclude that the presence of a charge gap and the absence of a spin
gap we find is consistent with the bosonization analysis.

In conclusion, we have presented a systematic study of the
non-universal exponent \( K_{\rho } \) in the Kondo lattice model, as
a function of the conduction electron density and the Kondo coupling.
The qualitative behavior of the charge structure factor $C(q)$ in the
weak and strong coupling limits could be ascribed to free
spin-$\frac{1}{2}$ and spinless fermions, respectively.  We also
showed that the spin correlation function can be described within a
Tomonaga-Luttinger liquid scheme only if logarithmic corrections are
included.  Finally, we have demonstrated that, although the charge
sector has a gap at quarter filling, there are gapless spin
excitations.

We thank A. Villares for useful discussions.
This work was supported by FAPESP 00/02802-7 (JCX), 01/00719-8
(JCX,EM), and CNPq 301222/97-5 (EM).

 \end{document}